\documentclass[twocolumn, prd, aps,superscriptaddress, preprintnumbers, tightenlines, showpacs, nofootinbib, eqsecnum, amsfonts, amsmath, floatfix]{revtex4-1}

\usepackage{epsfig}
\usepackage{graphics}
\usepackage{graphicx}
\usepackage{amsmath}
\usepackage{amssymb}
\usepackage{timestamp}
\usepackage{bm}
\usepackage[usenames,dvipsnames,svgnames,table]{xcolor}
\usepackage{xspace}
\usepackage{wasysym}
\usepackage{times}
\usepackage{appendix}
\usepackage{lipsum}
\usepackage[nolist,nohyperlinks]{acronym}
\usepackage{feynmp}

\def\be{\begin{equation}}
\def\ee{\end{equation}}
\def\ben{\begin{eqnarray}}
\def\een{\end{eqnarray}}
\def\nn{\nonumber}

\def\2p{{(2\pi)^2}}

\def\be{\begin{equation}}
\def\ee{\end{equation}}
\def\beq{\begin{equation}}
\def\eeq{\end{equation}}
\def\ben{\begin{eqnarray}}
\def\een{\end{eqnarray}}
\def\bes{\begin{subequations}}
\def\ees{\end{subequations}}

\def\nn{{\nonumber}}

\newcommand{\beqa}{\begin{eqnarray}}
\newcommand{\eeqa}{\end{eqnarray}}

\def\ikap0{{\cal J}_{\theta_0}(r)}

\def\one1{\langle \kappa_{(i)}\kappa_{(j)} \rangle}
\def\one{{[\bar \xi^{(ij)}]}}

\def\ba{\begin{eqnarray}}
\def\ea{\end{eqnarray}}

\def\2p{{(2\pi)^2}}

\def\be{\begin{equation}}
\def\ee{\end{equation}}

\def\beq{\begin{equation}}
\def\eeq{\end{equation}}

\def\ben{\begin{eqnarray}}
\def\een{\end{eqnarray}}

\def\nn{{\nonumber}}

\def\2p{{(2\pi)^2}}

\newcommand*{\diff}{\ensuremath{{\rm d}}}

\def\xd{\diff}

\usepackage{multirow}

\def\cb{\textcolor{black}}

\def\alan{\textcolor{black}}
\def\cL{\textcolor{black}}

\usepackage{colortbl}
\usepackage{color}

\def\cred{\textcolor{black}}

\usepackage[normalem]{ulem}

\def\abcd{\diff}

\SetSymbolFont{letters}{bold}{OML}{cmm}{b}{it}
\SetSymbolFont{operators}{bold}{OT1}{cmr}{bx}{n}

\usepackage{float}

\makeatletter

    \def\CT@@do@color{%
      \global\let\CT@do@color\relax
            \@tempdima\wd\z@
            \advance\@tempdima\@tempdimb
            \advance\@tempdima\@tempdimc
    \advance\@tempdimb\tabcolsep
    \advance\@tempdimc\tabcolsep
    \advance\@tempdima2\tabcolsep
            \kern-\@tempdimb
            \leaders\vrule
                    \hskip\@tempdima\@plus  1fill
            \kern-\@tempdimc
            \hskip-\wd\z@ \@plus -1fill }
    \makeatother

\definecolor{ZurichBlue}{rgb}{.255,.41,.884} 		
\definecolor{ZurichRed}{rgb}{0.9, 0.1, 0} 		
\definecolor{ZurichGreen}{rgb}{.196,.504,.396} 		
\definecolor{ZurichYellow}{rgb}{1,.648,0} 		
\definecolor{dodgerblue}{rgb}{0.12, 0.56, 1.0}
\definecolor{azure}{rgb}{0.0, 0.5, 1.0}
\definecolor{awesome}{rgb}{1.0, 0.13, 0.32}
\definecolor{alizarincrimson}{rgb}{0.82, 0.1, 0.26}
\definecolor{mediumpurple}{rgb}{0.58, 0.44, 0.86}
\definecolor{lasallegreen}{rgb}{0.03, 0.47, 0.19}
\usepackage[backref=page,colorlinks=true,linkcolor=dodgerblue,citecolor=dodgerblue,urlcolor=dodgerblue]{hyperref}

\begin{document}
\timestamp

\newcommand{\MSSL}{Mullard Space Science Laboratory, 
  University College London,\\
  Holmbury St Mary, Dorking, Surrey RH5 6NT, UK}
\newcommand{\Padova}{Dipartimento di Fisica e Astronomia ”G. Galilei”, Universit`a degli Studi di Padova, \\
  Via Marzolo 8, 35131 Padova, Italy }
\newcommand{\INFN}{INFN, Sezione di Padova, via Marzolo 8, I-35131, Padova, Italy}
\newcommand{\DMTP}{Department of Applied Mathematics and Theoretical Physics, University of Cambridge, \\
Wilberforce Road, Cambridge CB3 0WA, U.K.}
\newcommand{\Imp}{Imperial Centre for Inference and Cosmology (ICIC), Imperial College, London, SW7 2AZ,U.K.}
\title{Position-Dependent Correlation Function of Weak Lensing Convergence}
\author{D. Munshi}\affiliation{\MSSL}
\author{G. Jung}\affiliation{\Padova}\affiliation{\INFN}
\author{T. D. Kitching}\affiliation{\MSSL}
\author{J. McEwen}\affiliation{\MSSL}
\author{M. Liguori}\affiliation{\Padova}\affiliation{\INFN}
\author{T. Namikawa}\affiliation{\DMTP}
\author{A. Heavens}\affiliation{\Imp}
%
\begin{abstract}
%
  We provide a systematic study of the position-dependent correlation function in weak lensing
  convergence maps and its relation to the squeezed limit of the three-point correlation function (3PCF)
  using state-of-the-art numerical simulations. We relate the position-dependent correlation function
  to its harmonic counterpart, i.e., the position-dependent power spectrum
  or equivalently the integrated bispectrum.
We use a recently proposed improved fitting function, BiHalofit, for the bispectrum to compute the theoretical predictions
  as a function of source redshifts. In addition to low redshift results ($z_s=1.0-2.0$) we also
  provide results for maps inferred from lensing of the cosmic microwave background, i.e., $z_s=1100$.
  \cb{We include a {\em Euclid}-type realistic survey mask and noise. In
  agreement with the recent studies on the position-dependent power spectrum, we find 
  that the results from simulations are 
  consistent with the theoretical expectations when appropriate corrections are included.
  \cred{Performing a rough estimate, we find that the (S/N) for the detection of
    position-dependent correlation function 
    from {\em Euclid}-type mask with $f_{sky}=0.35$, 
    can range between $6-12$ depending on the value of the intrinsic ellipticity distribution parameter $\sigma_{\epsilon} = 0.3-1.0$.
    For reconstructed $\kappa$ maps using an ideal CMB survey the S/N $\approx 1.8$.
    We also found that a $10\%$ deviation in $\sigma_8$
    can be detected using IB for the optimistic case of $\sigma_\epsilon=0.3$
    with a S/N $\approx 5$. The S/N for such detection in case of $\Omega_M$
  is lower.}
  }

\end{abstract}
\maketitle

%
\section{Introduction}
\label{sec:intro}
%
%
Recently completed Cosmic Microwave Background (CMB) experiments, 
such as the \textit{Planck} Surveyor\footnote{{\href{http://http://sci.esa.int/planck/}{\tt Planck}}}\citep{Planck1}, 
have established a standard model of cosmology. 
Answers to many outstanding questions however remain unclear. These include, the nature of dark matter (DM)
and dark energy (DE), and possible modifications of General Relativity (GR) on cosmological scales \citep{MG1,MG2}.
In addition the sum of the neutrino masses \citep{nu} remains unknown. 
It is expected that the operational weak lensing surveys, including the
\alan{Subaru Hypersuprimecam survey}\footnote{\href{http://www.naoj.org/Projects/HSC/index.html}{\tt http://www.naoj.org/Projects/HSC/index.html}}(HSC)
\citep{HSC},
Dark Energy Survey\footnote{\href{https://www.darkenergysurvey.org/}{\tt https://www.darkenergysurvey.org/}}(DES)\citep{DES},
KiDS\citep{KIDS} and  near-future Stage-IV large scale structure (LSS)
surveys such as \textit{Euclid}\footnote{\href{http://sci.esa.int/euclid/}{\tt http://sci.esa.int/euclid/}}\citep{Euclid},
Rubin Observatory\footnote{\href{http://www.lsst.org/llst home.shtml}{\tt {http://www.lsst.org/llst home.shtml}}}\citep{LSSTTyson} and Roman Space Telescope\citep{WFIRST}, will provide answers to many of the questions that cosmology is facing
by directly probing the large-scale structure
and extracting information about clustering of the intervening  mass distribution in the Universe \citep{bernardeaureview}.
In contrast, spectroscopic galaxy redshift surveys such as
BOSS\footnote{\href{http://www.sdss3.org/surveys/boss.php}{\tt http://www.sdss3.org/surveys/boss.php}}\citep{SDSSIII}
or WiggleZ\footnote{\href{http://wigglez.swin.edu.au/}{\tt http://wigglez.swin.edu.au/}}\citep{WiggleZ} 
\cred{(also see Prime Focus Spectrograph\footnote{\href{http://pfs.ipmu.jp}{\tt http://pfs.ipmu.jp}} which is currently under development and the  Dark Energy Spectroscopic
  Instruments (DESI)\footnote{\href{http://desi.lbl.gov}{\tt http://desi.lbl.gov}} 
currently taking data)} probe
the distribution of galaxies as tracers and generally provide a biased picture \citep{biasreview}.

One challenge for weak lensing is that 
observations are sensitive to smaller scales where clustering is nonlinear and non-Gaussian \citep{bernardeaureview}, and are therefore 
difficult to model. A second challenge is that the statistical estimates of cosmological parameters based on power spectrum 
analysis are typically degenerate in particular
cosmological parameter combinations, e.g.\ $\sigma_8$ and $\Omega_{\rm M}$. To overcome these degeneracies external data sets (e.g.\ CMB), and the addition of 
tomographic or 3D \citep{3D} information are typically used. However, to address both of these challenges an alternative procedure
is to use higher-order statistics that probe the nonlinear regime \citep{higher1,higher2,higher3,MunshiBarber1,MunshiBarber2,MunshiBarber3}.

Gravitational clustering induces mode coupling that results in a secondary non-Gaussianity that
is more pronounced on smaller scales. This has led to development
of many estimators for the gravity-induced (secondary) non-Gaussianity from weak lensing surveys.
These statistics include the lower-order cumulants \citep{MunshiJain1} and their correlators \citep{MunshiBias},
the multispectra including the skew-spectrum \citep{AlanBi}, binned estimators \citep{binned1,binned2,binned3},
kurtosis spectra \citep{AlanTri}, Minkowski Functionals \citep{Minkowski}
as well as the entire PDF \citep{MunshiJain2}. Many of these estimators were initially
developed in the context of probing primordial non-Gaussianity \citep{Inflation}.
With a large fraction of sky-coverage, and the ability to detect a high number density of galaxies, surveys such as {\em Euclid} will be able to detect gravity-induced non-Gaussianity with a very
high signal-to-noise (S/N). In addition to lifting the degeneracy
in cosmological parameters, higher-order statistics are also
important for a better understanding of the covariance of lower-order estimators \citep{MunshiBarber4}.
The additional information content of the bispectrum, when added to that of the power spectrum, can significantly
\cred{reduce the errors in parameters \citep{Rizzato,Coulton,AnFengWang,SatoNihu, KayoTakadaJain1,TakadaJain1}
as well as provide a better} handle on systematics \citep{PyneJoachimi}.
In addition to the {\em summary statistics} and their estimators described above other approaches
of incorporating information regarding non-Gaussianity 
include likelihood based forward modelling\citep{Like} and likelihood-free techniques \citep{Likefree}.
In contrast to the derived statistics, these methods directly deal with the field variables but often
rely on expensive simulations or approximations to model gravitational dynamics.

\begin{figure}
  \includegraphics[width=0.35\textwidth]{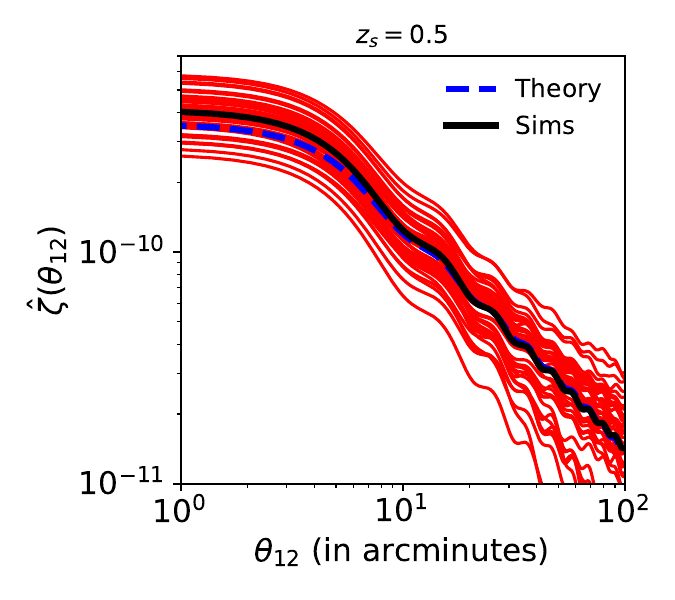}
  \vspace{-0.25cm}
  \caption{The sq3PCF $\zeta(\theta_{12})$ defined in Eq.(\ref{newdef:zeta_hat})  
    for the convergence map $\kappa$ for $z_s=0.5$ is shown as a function of $\theta_{12}$. The dashed line corresponds
    to the theoretical prediction.
    The thin solid lines correspond to the estimates from individual simulated
    convergence maps computed using 192 non-local patches described in section \ref{sec:ps}.
    The thick solid line represents the ensemble average of estimates from all maps. A total of 40 maps were used.}
   \label{fig:zs0d5}
\end{figure}

\begin{figure}
  \includegraphics[width=0.35\textwidth]{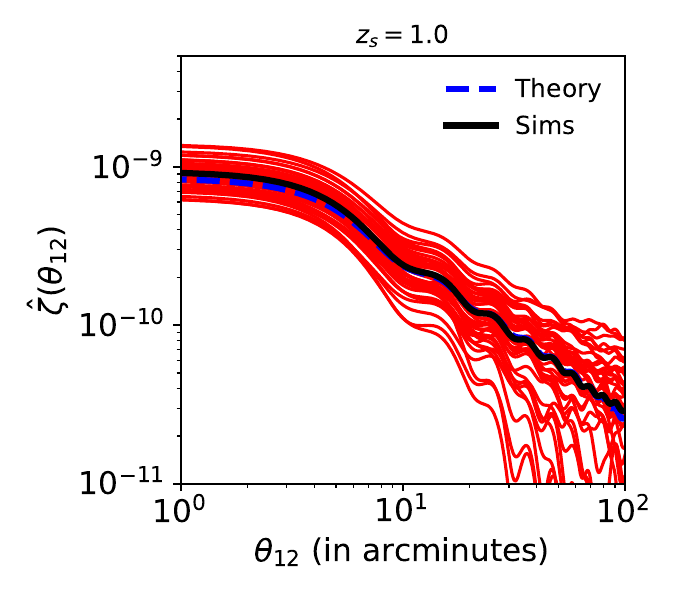}
  \vspace{-0.25cm}
   \caption{Same as Fig-\ref{fig:zs0d5} but for $z_s=1.0$}
  \label{fig:zs1d0}
\end{figure}

\begin{figure}
  \includegraphics[width=0.35\textwidth]{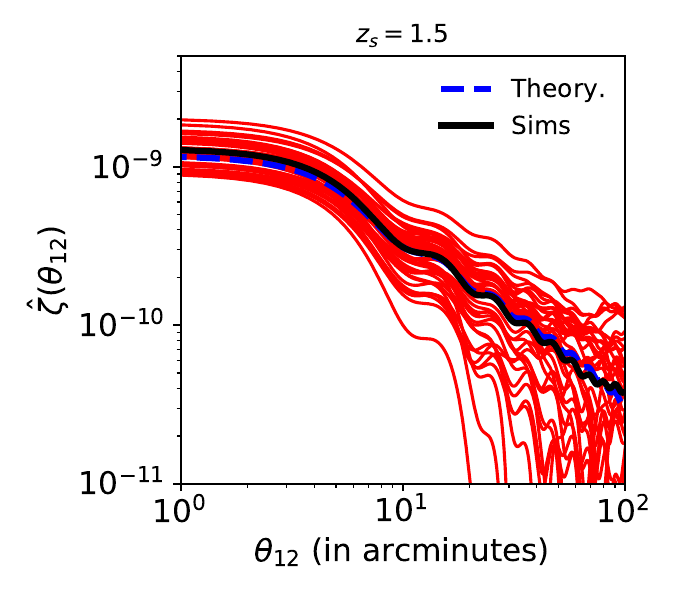}
  \vspace{-0.25cm}
  \caption{Same as Fig-\ref{fig:zs0d5} but for $z_s=1.5$}
  \label{fig:zs1d5}
\end{figure}

\begin{figure}
  \includegraphics[width=0.35\textwidth]{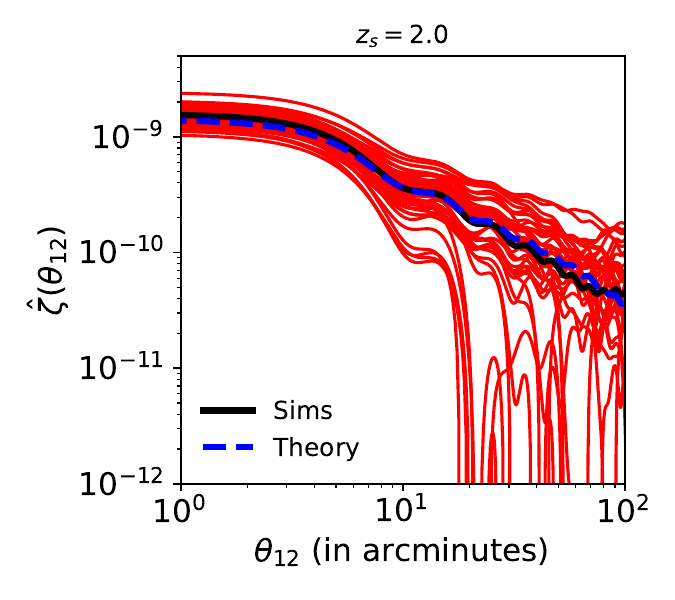}
  \vspace{-0.5cm}
  \caption{Same as Fig-\ref{fig:zs0d5} but for $z_s=2.0$}
  \label{fig:zs2d0}
\end{figure}

A complete characterization and estimation of bispectrum as well as its covariance can be demanding.
As a result, a subset of specific shapes of triangle that represent the bispectrum are usually considered.
Many recent papers have focussed on estimators that are particularly sensitive to the
squeezed configuration of the bispectrum known also as the {\em Integrated Bispectrum} (IB) \citep{PhDchiang}. These estimators
are particularly interesting because of their simplicity, as well as their ease of implementation \citep{Integrated,MGIB}.
In previous works such estimators have been used in 3D for quantifying galaxy clustering \citep{SDSSIII}, 21cm studies \citep{Giri},
the Cosmic Microwave Background (CMB) in 2D \citep{JungCMB}, as well as in 1D to probe Lyman-$\alpha$ absorption
features \citep{Lymanbispec,chiang2}. The IB estimator has also been applied to
weak lensing \citep{Jungwl}.
Our aim here is to develop these estimators for probing future 2D projected 
weak lensing surveys, and in particular {\em Euclid} \citep{Euclid}. Instead of focussing on the harmonic
domain \citep{Jungwl} we concentrate on the angular domain. Working in \alan{configuration space} 
has the advantage that the observational mask can be dealt with more easily than in harmonic space.
In this paper we will concentrate on the position-dependent two-point correlation function (2PCF) which probes the squeezed three-point correlation function (sq3PCF). 
This is complementary to its Fourier
counterpart, the position-dependent power spectrum, which on the other hand probes the squeezed configuration of
the bispectrum.
\begin{figure}
   \includegraphics[width=0.35\textwidth]{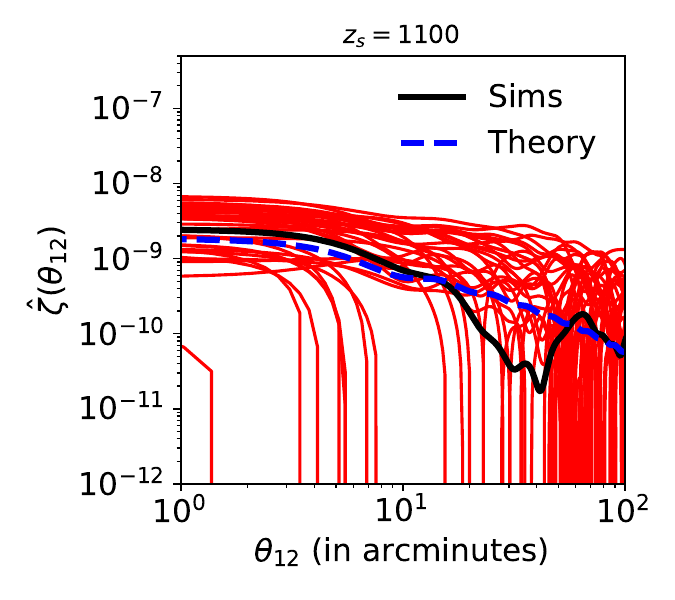}
  \vspace{-0.25cm}
  \caption{Same as Fig-\ref{fig:zs0d5} but for $z_s=1100$}
  \label{fig:zs11d0}
\end{figure}

This paper is organised as follows.  The introductory discussion
on weak lensing is presented in \textsection\ref{sec:wl}.
Some key results on position-dependent power spectrum are reviewed in \textsection\ref{sec:ps}.
Sec. \textsection\ref{sec:corr} introduces some of our key results.
The results of comparison against simulations are presented in \textsection\ref{sec:test}.
Finally the conclusions are drawn in \textsection\ref{sec:conclu}.
%
\section{Weak Lensing Three-Point Correlation Function}
\label{sec:wl}
%
The weak lensing convergence field $\kappa$ represents a line-of-sight integral of
the underlying matter density contrast $\delta$ between
the source plane at comoving distance $r_s$ (or redshift $z_s$) and the observer:
\begin{equation}
    \label{eq:convergence}
    \kappa({\boldsymbol \theta}, r_s) =  \int_0^{r_s} dr
    \, \omega(r, r_s)\delta({\boldsymbol \theta}, r)\,,
\end{equation}
In our notation, throughout,  ${\boldsymbol \theta}$ will represent the angular position on the
surface of the sky, and $r$ denotes the comoving distance.
The weight $\omega(r,r_s)$  appearing in the integral in Eq.(\ref{eq:convergence}) is given by
\begin{equation}
\label{eq:convergence-weight}
    \omega(r, r_s)  = \frac{3\Omega_M}{2}\frac{H_0^ 2}{c^2}\frac{d_A(r)d_A(r-r_s)}{a(r)d_A(r_s)}\,,
\end{equation}
where $d_A(r)$ denotes the comoving angular diameter distance, $a(r)$ is the scale factor,
and $\Omega_M$, $H_0$, $c$ represent the cosmological
matter density parameter, the Hubble constant and the speed of light, respectively. We have assumed a flat cosmology.
For reviews of weak lensing, see for example \citep{MunshiReview}.

We are mainly interested in the three-point correlator of the convergence field
and are thus concerned with the angle-averaged bispectrum denoted as $B_{\ell_1 \ell_2 \ell_3}$,
which can be constructed using the multipoles of $\kappa$ in the harmonic domain, $\kappa_{\ell m}$:
\ben
    \label{eq:bispectrum}
    && B_{\ell_1 \ell_2 \ell_3} = h_{\ell_1 \ell_2 \ell_3}\nonumber  \\
    && \times \sum\limits_{m_1 m_2 m_3} 
    \langle \kappa_{\ell_1 m_1} \kappa_{\ell_2 m_2} \kappa_{\ell_2 m_2}\rangle 
    \left ( \begin{array} { c c c }
     \ell_1 & \ell_2 & \ell_3 \\
     m_1 & m_2 & m_3
    \end{array} \right ),
\een
where the matrix is a Wigner $3j$-symbol,
and the geometrical factor $h_{\ell_1 \ell_2 \ell_3}$ is defined by
\begin{equation}
  \label{eq:h-factor}
    h_{\ell_1 \ell_2 \ell_3}
   \equiv
   \sqrt{\frac{(2\ell_1+1)(2\ell_2+1)(2\ell_3+1)}{4\pi}}
\left ( \begin{array} { c c c }
     \ell_1 & \ell_2 & \ell_3 \\
     0 & 0 & 0
\end{array} \right ).
\end{equation}
The power spectrum of $\kappa$ is defined as ${\cal C}_{\ell} \equiv \langle \kappa_{\ell m}\kappa^*_{\ell m} \rangle$.
The reduced bispectrum, used in the literature is defined as follows:
$b_{\ell_1 \ell_2 \ell_3} = B_{\ell_1 \ell_2 \ell_3} / h_{\ell_1 \ell_2 \ell_3}^2$.
In the Limber approximation, the bispectrum $B_{\ell_1 \ell_2 \ell_3}$
can be written in terms of the matter bispectrum
$B_\delta(k_1, k_2, k_3)$, where  $k_i$ are comoving wavenumbers, as
\ben
    \label{eq:convergence-bispectrum}
    && B_{\ell_1 \ell_2 \ell_3} =  h_{\ell_1 \ell_2 \ell_3}^2\int_0^{r_s} dr  \, \frac{\omega^3(r, r_s)}{d_A^4(r_s)} \nonumber \\
   && \quad\quad \times B_\delta\left(\frac{\ell_1}{d_A(r)}, \frac{\ell_2}{d_A(r)}, \frac{\ell_3}{d_A(r)};r\right)\,.
\een
Finally, to compute this, we  use the fitting function developed in  \cite{bispecfit}. We also incorporate
the post-Born correction \cite{PostBorn} which
introduces a significant contribution at high redshift but has a negligible effect at
low redshift \citep{Simulations}. We will discuss these issues 
in \textsection{\ref{sec:test}}.
%
\section{Position-dependent Power Spectrum}
\label{sec:ps}
The integrated bispectrum represents
the correlation of average of local convergence $\kappa_p$
estimated from a survey patch labelled by the index $p$ and the
local power spectrum estimated
from the same patch given by ${\cal C}_{\ell,p}$ (also called position-dependent power spectrum):

\begin{equation}
    \label{eq:ibisp-estimator}
    \hat{\cal B}_\ell = \frac{1}{N_\mathrm{p}}\sum\limits_{\mathrm{p}}
        \bar{\kappa}_\mathrm{p}^\mathrm{} {\cal C}_{\ell,\mathrm{p}}^\mathrm{}\,.
\end{equation}
Here, $N_p$ represents the total number of patches and by construction $\langle {\cal C}_{\ell,p} \rangle = {\cal C}_{\ell}$
and $\langle \bar\kappa_p \rangle = 0.$ Note that the patches can be localised in real space or in the
Fourier domain. To take into account the survey mask an elaborate procedure involving
Monte Carlo (MC) realisations exists in the literature \citep{JungCMB, Jungwl}. One of the advantages of working
in the real space, however, is that this can be circumvented. For the patches we have considered a {non-local} mask with band-limited
multipoles $w_{\ell m} = Y^*_{\ell m}({\boldsymbol \theta_0}); \; {\rm for} \;
\ell^w_{\rm min} \le \ell \le\ell^w_{\rm max}$.
Unlike local patches, which are typically used, our mask is non-zero for the entire sky.
We have chosen $\ell_{\rm min}=0$ and $\ell_{\rm max}=10$ for our study.
The centres of our patches $\boldsymbol \theta_0$ are chosen to be the centres of the
pixels at a HEALPix resolution of $N_{\rm side}=4$. Hence, for a given map we have a collection of $192$ patches.
We have chosen this to demonstrate the power of our method which can not be analysed with Limber-type approximation.

The following expression relates the integrated bispectrum $B_\ell$ with the
bispectrum $B_{\ell_1\ell_2\ell_3}$ introduced before \citep{JungCMB, Jungwl}:
\ben
    \label{eq:ibisp-bisp}
        {\cal B}_\ell = 
        && \frac{1}{N_\mathrm{p}} \frac{1}{4\pi g^2_{\rm sky} } \frac{1}{2\ell+1}
        \nonumber \\
        && \times \sum\limits_{\ell_1 \ell_2 \ell_3}
        \frac{B_{\ell_1 \ell_2 \ell_3}}{h_{\ell_1 \ell_2 \ell_3}}
        \sum\limits_{m_1 m_2 m_3} \begin{pmatrix}
            \ell_1 &  \ell_2 &  \ell_3\\
            m_1 & m_2 & m_3
        \end{pmatrix} \nonumber \\
        &&  \times \sum\limits_{m_4 m_5 m} (-1)^{m} 
        \begin{pmatrix}
            \ell &  \ell_1 &  \ell_4\\
            -m & m_1 & m_4
        \end{pmatrix} 
        \begin{pmatrix}
            \ell &  \ell_2 &  \ell_5\\
            m & m_2 & m_5
        \end{pmatrix} \nonumber \\
        && \times \sum\limits_{\mathrm{p}}(w^{\mathrm{p}}_{\ell_3 m_3})^{*}
        w_{\ell_4 m_4}^{\mathrm{p}} w^{\mathrm{p}}_{\ell_5 m_5}\,.
        \een
        Here, $g_{\rm sky}$ represents the fraction of sky coverage by individual patches.
        In contrast, the sky coverage of the entire survey is denoted by $f_{\rm sky}$. 
 The coefficients $w_{\ell m}$ denotes the harmonic multipoles of a given patch and $N_p$ is the number
 of patches considered. The above expression can be simplified for the type of patch we are using:
\ben
&& {\cal B}_\ell = 
 \frac{1}{4\pi g^2_{\rm sky} } \frac{1}{2\ell+1} \nonumber \\
&& \times \sum^{\ell+\ell_w^{max}}_{\ell_1\ell_2= \ell-\ell_w^{max}}\sum^{\ell_w^{\rm max}}_{\ell_3,\ell_4,\ell_5 ={\ell_w^{\rm min}}}
B_{\ell_1 \ell_2 \ell_3}  \mathcal{F}^{\ell \ell_1 \ell_2}_{\ell_3 \ell_4 \ell_5}.
\een

The following notation will be useful in simplifying expressions:
\ben
    \label{eq:factor-F}
     &&  \mathcal{F}^{\ell \ell_1 \ell_2}_{\ell_3 \ell_4 \ell_5} 
      =   (-1)^{\ell_2 + \ell_4} (2\ell_4+1)(2\ell_5+1) \nonumber  \\
     && \quad \times \left ( \begin{array} { c c c }
     \ell_1 & \ell_2 & \ell_3 \\
     0 & 0 & 0
      \end{array} \right )
      \left ( \begin{array} { c c c }
     \ell & \ell_1 & \ell_4 \\
     0 & 0 & 0
      \end{array} \right ) \nonumber \\
     && \quad
      \times \left ( \begin{array} { c c c }
     \ell & \ell_2 & \ell_5 \\
     0 & 0 & 0
      \end{array} \right )
      \left ( \begin{array} { c c c }
     \ell_3 & \ell_4 & \ell_5 \\
     0 & 0 & 0
      \end{array} \right ) \nonumber
      \left \{ \begin{array} { c c c }
     \ell_1 & \ell_2 & \ell_3 \\
     \ell_5 & \ell_4 & \ell
      \end{array} \right \}. \nonumber
      \een
The matrix in curly bracket denotes a $6j$ symbol.
The analytical expression for the covariance of IB, denoted as ${\mathbb C}_{\ell\ell'}$,
can be expressed in terms of the bispectrum covariance as follows:
\ben
    \label{eq:ibisp_covariance}
    && {\mathbb C}_{\ell\ell'} = (\delta{\cal B}_{\ell^{}}\delta{\cal B}_{\ell '})
        =  \frac{1}{(4\pi)^6 (g_\mathrm{sky})^4}
        \sum\limits_{\ell_{1,2,3,4,5}^{}} \sum\limits_{\ell'_{1,2,3,4,5}} \nonumber  \\
       && \quad\quad \mathrm \langle \delta B_{\ell_1^{} \ell_2^{} \ell_3^{}} \delta B_{\ell_1' \ell_2' \ell_3'}\rangle 
        \mathcal{F}^{\ell^{} \ell_1^{} \ell_2^{}}_{\ell_3^{} \ell_4^{} \ell_5^{}} 
       \mathcal{F}^{\ell' \ell'_1 \ell'_2}_{\ell'_3 \ell'_4 \ell'_5}.
       \een    
       For a noise-dominated case the bispectrum covariance can be approximated by the following
       Gaussian expression.
\ben
   \label{eq:bispectrum-covariance}
   && (\delta B_{\ell_1^{} \ell_2^{} \ell_3^{}}\delta B_{\ell_1' \ell_2' \ell_3'}) =
   h_{\ell_1^{} \ell_2^{} \ell_3^{}}^2 {\cal C}_{\ell_1^{}}{\cal C}_{\ell_2^{}}{\cal C}_{\ell_3^{}}
   \nonumber \\
   && \quad \times\, ( \delta_{\ell_1^{} \ell_1'}\delta_{\ell_2^{} \ell_2'}\delta_{\ell_3^{} \ell_3'} 
    + \mathrm{cyc. per.} ).
\een
The ${\cal C}_{\ell}$ in this case takes contributions from both signal and noise:
\ben
    \label{eq:ibisp_variance}
      &&  {\mathbb C}_{\ell\ell} \simeq \frac{1}{(4\pi)^6 g^4_{sky}} 
               \sum\limits_{\ell_{1,2}^{} = \ell - \ell_max }^{\ell + \ell_wmax } 
                \sum\limits_{\ell_{3,4,5}^{} = \ell_min } ^{\ell_w}
                \sum\limits_{\ell'_{4,5} = \ell_min } ^{\ell_w} \nonumber \\
     && \times h_{\ell_1^{} \ell_2^{} \ell_3^{}}^2  {\cal C}_{\ell_1^{}}{\cal C}_{\ell_2^{}}{\cal C}_{\ell_3^{}} 
       \mathcal{F}^{\ell^{} \ell_1^{} \ell_2^{}}_{\ell_3^{} \ell_4^{} \ell_5^{}}
       (\mathcal{F}^{\ell^{} \ell_1^{} \ell_2^{}}_{\ell_3^{} \ell'_4 \ell'_5}
   + \mathcal{F}^{\ell^{} \ell_2^{} \ell_1^{}}_{\ell_3^{} \ell'_4 \ell'_5} ) \,.
\een
We have focussed on a nonlocal patch of a sky. This requires
the all-sky formalism presented above.

\begin{figure}
  \includegraphics[width=0.35\textwidth]{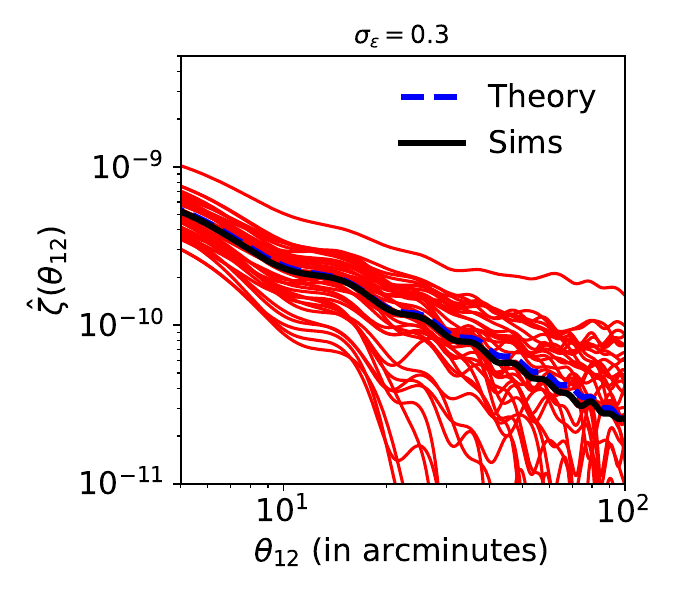}
  \caption{Same as Fig-\ref{fig:zs0d5} but with an {\em Euclid} like mask and noise included.
    We use a pseudo Euclid mask  which removes both the galactic and elliptic planes. The resulting fraction of
    sky coverage is $f_{\rm sky} = 0.35$ (see \citep{Minkowski} for more details).
    We also assumea  Gaussian noise, with a noise power spectrum amplitude given by $n_\ell = \sigma^2_{\epsilon}/ \bar{\rm N}$.
    We have taken $\bar {\rm N}=30\, \rm arcmin^{-2}$ as expected for {\em Euclid}. We have also taken $\sigma =0.3$. The sources
  are placed at a redshift $z_s=1$.}
  \label{fig:sigmaA}
\end{figure}

\begin{figure}
  \includegraphics[width=0.35\textwidth]{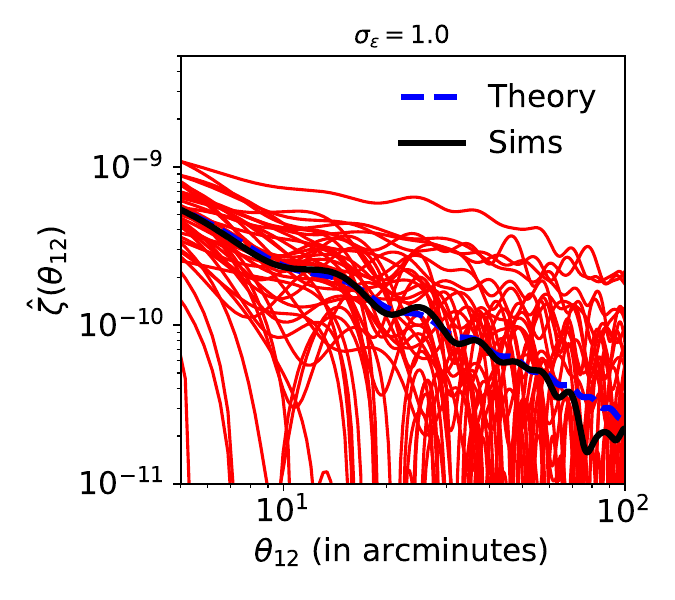}
  \caption{Same as Fig.\ -\ref{fig:sigmaA}  but for $\sigma_{\epsilon} =1.0$.}
  \label{fig:sigmaB}
\end{figure}
%
\section{Position-dependent Correlation Function}
\label{sec:corr}
%
\begin{table}
\begin{center}
\begin{tabular}{ |c |c|c|c|c|c| } 
\hline
 & 3\' & 10\' &30\' & 100\'& S/N \\
\hline
$\sigma_{\epsilon}=0.3$ & 5.12 & 3.61 & 2.10 & 0.73 &11.578  \\
\hline
 $\sigma_{\epsilon}=1.0$ & 3.27 & 1.73 & 0.79  & 0.43 & 6.241 \\ 
\hline
\end{tabular}
\end{center}
\caption{The $\zeta / \sigma(\zeta) $
  is presented for various separation annles.
  The top row corresponds to $\sigma_\epsilon=0.3$ and the botton row corresponds to  $\sigma_\epsilon=1.0$.
  The resulting total $\rm S/N$ is also presented.}
\label{zlable1}
\end{table}
\begin{table}
\begin{center}
\begin{tabular}{ |c|c|c|c|c|c| } 
\hline
   3\' & 10\' &30' & 100\'& S/N \\
\hline
  0.944 & 0.432 & 0.251 &  0.145 &1.772 \\
\hline
\end{tabular}
\end{center}
\caption{Same as Table-\ref{zlable1}, but for $z_s=1100$.}
\label{table2}
\end{table}


The implementation of the position-dependent power spectrum,
or equivalently the Integrated Bispectrum, was presented recently in an accompanying paper \citep{Jungwl}.
However, for smaller sky coverage, the real space analogue of the position-dependent power spectrum,
namely the position-dependent correlation function, defined in real space can be equally useful.
Our aim in this paper is to express the squeezed limit of the three-point correlation function or 3PCF, 
in terms of the position-dependent correlation function (see \citep{SDSSIII}
for equivalent derivation in 3D for galaxy clustering statistics). We will show how this can be related
to the response function approach. We will also relate these results in the position space
with the ones found in harmonic domain. We will show  how the squeezed 3PCF and the squeezed bispectrum
are related in 2D. To this aim, we will start by defining the global correlation function $\xi(\theta_{12})$.
Though the results are primarily derived  keeping the projected weak-lensing convergence maps in mind
they are of generic nature and will be valid for any projected field.
The global correlation function $\xi(\theta_{12})$ in a 2D projected sky covering area $A_s$ for convergence $\kappa$
is defined through the following expression:
\bes
\ben
&&\xi({\theta}_{12}) \equiv \langle\kappa({\bm\theta})\kappa({\bm\theta}+{\bm\theta}_{12}) \rangle; \nonumber \\
&& \quad\quad = {1 \over A_s} \int {\abcd\,\varphi_{12} \over 2\pi} \int \abcd^2{\bm\theta}\;\;
\langle \kappa({\bm\theta})\kappa({\bm\theta}+{\bm\theta}_{12})\rangle.
\label{eq:global}
\een
\ees
Here, $\varphi_{12}$ denotes the polar angle associated with the vector ${\boldsymbol\theta}_{12}$.
The assumption of isotropy and homogeneity allows us to write $\xi(\theta_{12})$ as a function of
separation $\theta_{12}$ and not its orientation. Notice for the global correlation the points
$\bm\theta$ and $\bm\theta+\bm\theta_{12}$ are both assumed to be within the survey patch.
The local estimate is estimated within a patch of the sky of area $A_p$:
\ben
\hat \xi(\theta_{12}) = {1\over A_p}\int {{\xd\,\varphi_{12} \over 2\pi}}
\int \xd^2\,{\boldsymbol \theta}\, \kappa({\bm\theta+\bm\theta_{12}})\kappa({\bm\theta}).
\label{eq:local}
\een
Using a 2D window function $W$ the local correlation function $\hat\xi({\theta}_{12})$ can be expressed as:
\ben
&& \hat\xi({\theta}_{12}) = {1\over A_s}\int{\xd\,\varphi_{12}\over 2\pi}
\int \xd^2{\boldsymbol\theta}\; \langle\kappa({\boldsymbol\theta}+{{\boldsymbol\theta}_{12}})
\kappa(\boldsymbol\theta_{})\rangle \nn \\
&& \; \quad\quad \times W(\boldsymbol\theta+\boldsymbol\theta_{12})W({\boldsymbol\theta}_{}).
\een
Indeed, it can be easily shown from Eq.(\ref{eq:local}) that $\hat\xi(\theta_{12})$ is not an unbiased estimator of the global $\xi(\theta_{12})$.
We introduce the multiplicative bias factor $s(\theta_{12})$ to relate the two,
i.e.\ $\langle\hat\xi(\theta_{12})\rangle=s(\theta_{12})\xi(\theta_{12})$ which depends
on the survey geometry. The multiplicative factor $s(\theta_{12})$, which originates from the \alan{finite-volume correction,}
is as follows:
\ben
&& \langle\hat\xi(\theta_{12})\rangle=s(\theta_{12})\xi(\theta_{12}); \nn \\
&& \quad s(\theta_{12}) \equiv{1\over A_s}\int{\xd\varphi_{12} \over 2\pi}\int {\xd^2{\boldsymbol\theta}}
\; W(\boldsymbol{\theta}+{\bm\theta}_{12})W(\boldsymbol\theta).
\label{eq:s12}
\een
By cross-correlating the local estimates of 2PCF $\hat\xi$ and the mean $\bar\kappa$ from the same patch we
arrive at the following estimate for the sq3PCF denoted as $\zeta({\theta}_{12})$:
\ben
&& \zeta_{}(\theta_{12}) \equiv \langle\hat\xi({\theta}_{12}){\bar\kappa}\rangle \nn  \\
&& = {1\over A_s^2}\int {\xd\varphi_{12} \over 2\pi}
\int \xd^2\boldsymbol\theta_{1} \int \xd^2\boldsymbol\theta_2\;
   \zeta(\boldsymbol\theta_1+\boldsymbol\theta_{12},\boldsymbol\theta_1,\boldsymbol\theta_2) \nn \\
   && \hspace{2cm} \times W(\boldsymbol\theta_1+\boldsymbol\theta_{12})W({\boldsymbol\theta}_1)
   W({\boldsymbol\theta}_2). 
   \label{eq:IBdefine}
\een
An unbiased estimator independent of survey geometry can be constructed using the following expression:
\ben
\hat \zeta(\theta) = {1 \over s(\theta)} \zeta(\theta).
\label{newdef:zeta_hat}
\een
In the response function approach we expand the estimated two-point correlation function as a function
of $\bar\kappa$: $\hat{\xi}(\theta) = \xi(\theta)|_{{\bar\kappa}=0} + {d \xi/d \bar\kappa}{\big |}_{\bar \kappa=0} \bar\kappa + \cdots$.
On cross-correlating with $\bar\kappa$, at the lowest order we get the squezeed limit of the 3PCF introduced
above: $\zeta(\theta) \equiv \langle{\bar\kappa}\hat{\xi}(\theta) \rangle = {d \xi/d \bar\kappa}{\big |}_{\bar \kappa=0} \langle{\bar\kappa}^2\rangle $.
The integrated bispectrum and the integrated 3PCF are related through the following expression:
\ben
\zeta({\bf\theta}_{12}) = {1 \over 4\pi} \sum_{\ell} (2\ell+1) P_{\ell}(\cos \theta_{12}) {\cal B}_{\ell}.
\een
Here $P_{\ell}$ is the Legendre polynomial of order $\ell$.
The new observable introduced above is easy to interpret and can
be estimated using tools developed
for estimation of two-point statistics $\xi$, thus side-stepping the
complexeity involved in estimation
of three-point statistics. Evaluation of two-point correlation
function from cosmological data sets has a rich history and
many different estimators exist which can be exploited to compute the
sq3PCF.
%
%
\section{Comparison against Simulations}
\label{sec:test}
%
\begin{table}
  \begin{center}
\begin{tabular}{ |c|c|c|c|c|c| } 
\hline
 & 3\' & 10\' &30\' & 100\'& S/N \\
\hline
$\sigma_{\epsilon}=1.0$ & 0.14/0.12 & 0.08/0.01 & 0.05/0.07 & 0.04/0.06 & 0.24/0.26  \\
\hline
 $\sigma_{\epsilon}=0.3$ & 0.09/0.08 & 0.00/0.01 & 0.01/0.03 & 0.02/0.03 & 0.013/0.14 \\ 
\hline
\end{tabular}
  \end{center}
\caption{The $|\delta\zeta| / \sigma(\zeta) $
  is presented for various separation angles.
 The top row corresponds to $\sigma_\epsilon=0.3$
 and the botton row corresponds to  $\sigma_\epsilon=1.0$.
 The resulting total $\rm S/N$ is also presented.
 Two entries for a given $\theta_{12}$ and $\sigma_\epsilon$
 correspond to $\delta \zeta = \zeta_{\Omega+}- \zeta_{\Omega}$
 and $\delta\zeta_{-} = \zeta_{\Omega-}- \zeta_{\Omega} $.
 The quantites  $\zeta_{\Omega+}$ and  $\zeta_{\Omega-}$ are computed using
 10\% higher and lower values of $\Omega_M$.}
\label{tab:s2n1}
\end{table}

\begin{table}
\begin{center}
\begin{tabular}{ |c|c|c|c|c|c|c| } 
\hline
 &  3\' & 10\' &30\' & 100\'& S/N \\
\hline
  $\sigma_{\epsilon}=1.0$ & 0.09/0.08 & 0.00/0.00 & 0.02/0.03 & 0.02/0.03 & 0.02/0.62  \\
\hline
 $\sigma_{\epsilon}=0.3$ & 2.57/1.19 & 1.75/1.31 & 0.09/0.71 & 0.31/0.84 & 5.57/4.77 \\ 
\hline
\end{tabular}
\end{center}
\label{Tab:diff2}
\caption{Sama as Table-\ref{tab:s2n1} but the entries correspond to 10\% higher and
  lower values of the paramter $\sigma_8$.}
\label{tab:s2n2}
\end{table}

The computation of the sq3PCF $\zeta$ relies on
estimation of the 2PCF $\xi$. Thus its implementation is rather simple
and computationally inexpensive. The computation of $\zeta$
is done by dividing the maps into many different patches. 
We have focussed on non-local patches
that are non-zero on the entire celestial sphere and
can only be analysed using an all-sky approach as this
approach is based on spherical harmonics. 
The two-point correlation function
for the convergence maps $\kappa$ is sensitive to small
scale modes, when correlated with the average $\bar\kappa$ estimated
from the same patch can give an estimate of the 3PCF
in the squeezed limit, i.e. $\zeta$. The resulting estimator corresponds to the
estimator described in Eq.(\ref{eq:IBdefine}).
The choice of patches can have a high impact on the signal-to-noise
of the estimated sq3PCF. In addition to patches that we considered
here, other filtering functions or non-local patches
can be considered.

Unlike the position-dependent power spectrum, where
spurious bispectral modes are induced by the mask, 
requiring an elaborate Monte-Carlo-based
subtraction procedure \citep{JungCMB,Jungwl},
the position-dependent two-point correlation function
is free from such complications. We have used two different techniques
in our estimation of $\zeta$. First we have used the position-dependent power spectra
from \citep{Jungwl} and used it to reconstruct the $\zeta$. We have
also used a publicly available software TreeCorr\footnote{https://github.com/rmjarvis/TreeCorr}\citep{treecorr}
to directly estimate the two-point correlation function to check the results though the results from TreeCorr are not shown. 


The state-of-the art simulations that we use are presented in \citep{Simulations}.
\footnote{http://cosmo.phys.hirosaki-u. ac.jp/takahasi/allsky\_raytracing/nres12.html.}
We use the convergence maps at HEALPix\footnote{http://healpix.sourceforge.net}\citep{HEALPix} resolution
$N_{\rm side} = 4096$ and downgrade their resolution to $N_{\rm side} = 2048$ and use
$\ell_{\rm max} = 2000$ for all the analyses presented here.


Traditionally patches that are localised in real space
are considered by dividing the map in smaller patches. In this approach a Limber approximation
can be used to simplify the analytical results.
In this study, we have considered patches that are localised in the harmonic domain.
The estimates from individual maps are the averages of all patches constructed from that map.
We have used a total 40 maps and from each of these maps we constructed 192 patches
(see \textsection\ref{sec:ps} for more details on construction of these non-local patches).


Our theoretical and numerical results are
presented in Fig.\ \ref{fig:zs0d5}, Fig.\ \ref{fig:zs1d0},  Fig.\ \ref{fig:zs1d5} and  Fig.\ \ref{fig:zs2d0}
respectively for redshifts $z_s=0.5,1.0,1.5$ and $2.0$. 
The thick solid lines in each of these panels correspond to the ensemble average of all maps
and the thin solid lines for individual maps.
\cb{The results for an individual map represent the average of all the patches constructed from that map.
The theoretical expectation are shown as dashed lines.}


We have also considered realisations of $\kappa$ maps inferred from CMB observations.
Our results for all redshifts include the post-Born corrections \citep{PostBorn}.
The post-Born corrections to the 3PCF are included in our modelling of $\zeta(\theta)$,
although such corrections do not contribute significantly at lower redshifts \citep{SkewSpec,Minkowski}.
The lensing signal for $z_s=1100$ is rather weak, but we get reasonable results for small angular scales.
For large angular scales the recovered $\zeta$ shows large fluctuations. The results
are presented in Fig.\ \ref{fig:zs11d0}.


In addition to the noise-free all-sky simulations we have also
considered maps at source redshift $z_s=1.0$ and applied the
pseudo-Euclid mask (see \citep{SkewSpec,Minkowski} for a detailed description).
This mask removes both the galactic and elliptic planes thus leaving roughly $f_{\rm sky} = 0.35$
for science exploitation.  We also add two different levels of Gaussian noise.
The noise power spectrum denoted as $n_{\ell}$ for the noise is given by $n_{\ell} = {\sigma^2_{\epsilon}/ \bar N}$.
For Euclid we have taken $\bar n = 30\;{\rm arcmin}^{-2}$.
We have considered two values for $\sigma_{\epsilon}$.
The results for $\sigma_{\epsilon}=0.3$ are shown in (Fig.\ \ref{fig:sigmaA}) and for $\sigma_{\epsilon}=1.0$.
in (Fig.\ \ref{fig:sigmaB}).
As expected the results of comparison of theoretical and numerical results are in agreement with
noise free case. The addition of noise only increases the scatter.
%
\section{Conclusion}
\label{sec:conclu}
%
Using an estimator designed to probe sq3PCF and state-of-the-art simulations we found that the
analytical results can be very accurately recovered from numerical simulation.
Our estimator probes the squeezed configuration of the bispectrum.

In previous studies, using a different but related estimator, known also as the binned
estimator, it was found \citep{Jungwl} that, for other shapes,
\alan{including e.g. the equilateral shape}, the
fitting function we have used,  provides rather accurate description of the
numerical estimates from simulations. However, for squeezed configurations
this was not the case. \cL{We have presented the corresponding results of analysis in the configuration space
in this paper.}

\alan{For cosmological parameter inference using the position-dependent
correlation function or IB, if we make the further assumption that the likelihood has gaussian form, we still 
require an accurate 
covariance matrix, and this is a far from trivial issue.}
Most formalism borrowed from CMB studies use a Gaussian Likelihood
 or its variants. \cred{We notice that recently
it was shown that a Gaussian approximation is sufficient
for the power spectrum \citep{Upham} and one-point
third-order moment \citep{PSimon} for the aperture mass
$\langle M^3_{ap}\rangle$. However, similar
study for two-point third-order statistics, i.e., $\zeta(\theta_{12})$
or equivalently ${\cal B}_{\ell}$ in the harmonic domain, is currently lacking
in the literature.}

While the diagonal entries in the covariance matrix can
be modelled numerically using relatively few simulations,
accurate numerical estimates for the off-diagonal elements
require  many more realisations than we have currently available.

\cred{We have computed the scatter in $\zeta$ represented as $\zeta$
and tabulated $\zeta/\sigma(\zeta)$ in Table-I and Table -II
as a function of the separation angle $\theta_{12}$. Two values
of $\sigma_{\epsilon}=0.3$ and $1.0$ are considered which correspond to
the (S/N) of $11.6$ and $6.24$ respectively. A {\em Euclid}-type mask with $f_{sky}$=0.3
was considered and sourced were placed at $z_s=1.0$.
For $z_s=1100$ we get ${\rm (S/N)} = 1.8$.}

\cred{We have also studied the  
  sensitivity of $\zeta$ to cosmological parameters.
  We have computed $|\delta\zeta| / \sigma(\zeta) $.
  This is done by constructing 
  $|\delta\zeta| = \zeta_{\Omega+}- \zeta_{\Omega}$
 and $|\delta\zeta| = \zeta_{\Omega-}- \zeta_{\Omega} $ as a function of
  $\theta_{12}$ and $\sigma_\epsilon$. Here, $\zeta_{\Omega+}$ and
 $\zeta_{\Omega-}$  corrrespond respectively to 10\% higher and lower value
 of $\Omega_M$. The results are presented in Table-\ref{tab:s2n1} for $\Omega_M$.
 Corresponding results for $\sigma_8$ are shown in Table-\ref{tab:s2n2}
 The top row corresponds to intrinsic ellipticity distribution
 parameter $\sigma_\epsilon=1.0$
  whereas the bottom row corresponds to  $\sigma_\epsilon=0.3$.
 The total $\rm (S/N)$,
  for 10\% deviation in 
  can be as high as $5.5$ for $\sigma_8$.}

At the time of writing this paper, we found a similar study \citep{Halder} in which the authors consider
the squeezed 3PCF for shear using compensated filter and compare the results from simulation against analytic prediction
using an older fitting function for the matter bisepctrum.


\section*{Acknowledgment}
DM is supported by a grant from the Leverhume Trust at MSSL.
DM would like to thank Peter Taylor, Ryuichi Takahashi for many useful discussions. 
\bibliography{corr.bbl}
\end{document}